\documentclass[12pt]{article}
\usepackage{amsmath, amssymb,graphicx}

\newwrite\ffile\global\newcount\figno \global\figno=1

\def\writedef#1{}

\input epsf
\def\figin{\epsfcheck\figin}\def\figins{\epsfcheck\figins}
\def\epsfcheck{\ifx\epsfbox\UnDeFiNeD
\message{(NO epsf.tex, FIGURES WILL BE IGNORED)}
\gdef\figin##1{\vskip2in}\gdef\figins##1{\hskip.5in}% blank space instead
\else\message{(FIGURES WILL BE INCLUDED)}%
\gdef\figin##1{##1}\gdef\figins##1{##1}\fi}

\def\figinsert{}
\def\ifig#1#2#3{\xdef#1{fig.~\the\figno}
\writedef{#1\leftbracket fig.\noexpand~\the\figno}%
\figinsert\figin{\centerline{#3}}\medskip\centerline{\vbox{\baselineskip12pt
\advance\hsize by -1truein\center\footnotesize{  Fig.~\the\figno.} #2}}
\bigskip\endinsert\global\advance\figno by1}
\def\endinsert{}

\usepackage{amssymb}
\begin{document}
\baselineskip 18pt
\newcommand{\Tr}{\mbox{Tr\,}}
\newcommand{\beq}{\begin{equation}}
\newcommand{\eeq}{\end{equation}}
\newcommand{\bea}{\begin{eqnarray}}
\newcommand{\eea}[1]{\label{#1}\end{eqnarray}}
\renewcommand{\Re}{\mbox{Re}\,}
\renewcommand{\Im}{\mbox{Im}\,}

\def\N{{\cal N}}

%%%%%%
%%%%%%%%%%%%%%%%   TITLE    %%%%%%%%%%%%%%%%%%%%

\thispagestyle{empty}
\renewcommand{\thefootnote}{\fnsymbol{footnote}}

{\hfill \parbox{4cm}{
        HU-EP-03-27 \\
        SHEP-03-10 \\
}}

\bigskip

\begin{center} \noindent \Large \bf
Chiral Symmetry Breaking  and Pions in

{ \noindent \Large \bf Non-Supersymmetric Gauge/Gravity Duals}
\end{center}

\bigskip\bigskip\bigskip

\centerline{ \normalsize \bf J.~Babington $^{a}$, J.~Erdmenger
$^a$, N.~Evans $^b$,
Z.~Guralnik $^{a}$ and
I.~Kirsch $^{a}$\footnote[1]{\noindent \tt
james@physik.hu-berlin.de, jke@physik.hu-berlin.de, evans@phys.soton.ac.uk,
zack@physik.hu-berlin.de,  ik@physik.hu-berlin.de} }

\bigskip
\bigskip\bigskip

\centerline{$^a$ \it Institut f\"ur Physik}
\centerline{\it Humboldt-Universit\"at zu Berlin}
\centerline{\it Newtonstra\ss e 15 }
\centerline{\it D-12489 Berlin, Germany}
\bigskip\bigskip

\centerline{$^b$ \it Department of Physics}
\centerline{ \it Southampton University}
\centerline{\it  Southampton, S017 1BJ }
\centerline{ \it United Kingdom}
\bigskip

\bigskip\bigskip

\renewcommand{\thefootnote}{\arabic{footnote}}

\centerline{\bf \small Abstract}
\medskip

{\small \noindent We study gravity duals of large $N$
non-supersymmetric gauge theories with matter in the fundamental
representation by introducing a D7-brane probe into deformed AdS
backgrounds. In particular, we consider a D7-brane probe in both the AdS
Schwarzschild black hole solution and in the background found by
Constable and Myers, which involves a non-constant dilaton and
$S^5$ radius. Both these backgrounds exhibit confinement of 
fundamental matter and a
discrete glueball and meson spectrum.   We numerically compute the
$\bar\Psi\Psi$ condensate and meson spectrum associated with these
backgrounds. In the AdS-black hole background, a quark-bilinear
condensate develops only at non-zero quark mass. We speculate on
the existence of a  third order phase transition at a critical
quark mass where the D7 embedding undergoes a geometric
transition. In the Constable-Myers background, we find a chiral
symmetry breaking condensate as well as the associated Goldstone
boson in the limit of small quark mass.  The existence of the
condensate ensures that the D7-brane never reaches 
the naked singularity at the origin of the deformed $AdS$ space.}

\newpage

%%%%%%%%%%%%%%%%%%%%%%%%%%%%%%%%%%%%%%%%%%%%%%%%%%%%%%%%%

\section{Introduction}

The discovery of the AdS/CFT correspondence \cite{Mal,Gub,Wit} has
led to promising new ideas for studying strong coupling phenomena
in large $N$ gauge theories. Generalizations of the correspondence
in which conformal symmetry and supersymmetry are broken are
potentially useful for describing realistic quantum field
theories. In particular, it is hoped that methods based on
gauge-gravity duality will eventually be applicable to QCD. The
simplest generalizations involve deforming AdS by the
inclusion of relevant operators \cite{Girardello:1998pd}. 
These geometries are
asymptotically AdS, with the deformations interpreted as
renormalization group (RG) flow from a super-conformal gauge
theory in the ultraviolet to a QCD-like theory in the infrared. Moreover
a number of non-supersymmetric ten-dimensional geometries of this
or related form have been found
\cite{Witten,Gubser:1999pk,Babington:2002ci,Babington:2002qt,CM} and
have been shown to
describe confining gauge dynamics. There have been
interesting calculations of the glueball spectrum in three and
four-dimensional QCD by solving classical supergravity equations
in various deformed AdS geometries
\cite{Csaki,Koch,Zyskin,Russo,Minahan,Ooguri,Csaki2,Russo2,Csaki3,Constable2,Brower}.

A difficulty with describing QCD in this way arises due to the
asymptotic freedom of QCD. The vanishing of the 't~Hooft coupling
in the UV requires the dual geometry to be infinitely
curved in the region corresponding to the UV. In this case
classical supergravity is insufficient and one needs to use full
string theory. Formulating string theory in the relevant
backgrounds has thus far proven difficult.  The existing glueball
calculations involve geometries with small curvature that return
asymptotically to AdS (the field theory returns to the strongly
coupled ${\cal N}=4$ theory in the UV), and are in the same
coupling regime as strong coupling lattice calculations far from the
continuum limit. There is nevertheless optimism that the glueball
calculations are fairly accurate, based on comparisons with
lattice data \cite{Teper,Morningstar,CE}.

In this paper we shall discuss progress on a further problem which
must be solved to study QCD using gravity, namely the inclusion of
particles in the fundamental representation of the gauge group,
i.e.~quarks. Since AdS geometries
arise as near-horizon limits for coincident branes, the dual field
theories have only adjoint matter.  The same is true for
deformations of AdS. It turns out that fundamental representations
can be introduced by including appropriately embedded probe
branes. Examples of this  \cite{KarchRandall,DFO,EGK,CEGK1}
were found by embedding a probe brane on an $AdS_d$ subspace of
the full $AdS_D$ geometry,  where $d<D$ and the boundary of
$AdS_d$ is part of the boundary of the $AdS_D$. Such embeddings
give rise to a dual field theory with ``quarks" that are not free
to move in all spatial directions. These ``defect'' conformal
field theories are interesting for a variety of reasons, for instance
for localizing gravity \cite{KarchRandall} or for their mathematical properties
\cite{Constable:2002vt}, but are somewhat far from QCD.

In order to obtain fundamental fields in four space-time dimensions,
Karch and Katz \cite{KarchKatz,weiner} consider a
configuration in which a D7-brane in $AdS_5 \times S^5$ fill the
$AdS_5$ and wrap an $S^3$ inside $S^5$. This configuration is dual
to a four-dimensional ${\mathcal N}=2$ Yang-Mills theory
describing open strings in the presence of one D7 and $N$
D3-branes sharing three spatial directions. The degrees of freedom
are those of the ${\mathcal N} =4$ super Yang-Mills theory,
coupled to an ${\mathcal N} =2$ hypermultiplet with fields in the
fundamental representation of SU(N). The latter arise from strings
stretched between the D7 and D3-branes.  When the D7 and D3's are
separated, the fundamental matter becomes massive  and the dual
description involves a probe D7 on which the induced metric is
only asymptotically $AdS_5 \times S^3$.  In this case there is a
discrete spectrum of mesons.  This spectrum has been computed
(exactly!) at large 't~Hooft coupling \cite{MateosMyers} using an
approach analogous to the glueball calculations in deformed AdS
backgrounds. The novel feature here is that the ``quark'' bound
states are described by the scalar fields in the Dirac-Born-Infeld
action of the D7 brane probe.

In view of a gravity description of Yang-Mills theory with confined quarks,
it is natural to attempt to generalize these calculations to
probes of deformed AdS spaces. For instance in \cite{Sonnenschein}, 
a way to embed D7-branes in the
Klebanov-Strassler background \cite{KS} was found, following the suggestion of
\cite{KarchKatz}.   Moreover in \cite{Sonnenschein}, the
spectrum of mesons dual to fluctuations
of the D7-brane probe in the KS-geometry was calculated.
The underlying theory is an
${\mathcal N}=1$ gauge theory with massive chiral superfields in
the fundamental representation. - Related work may also be found in
\cite{Nastase:2003dd}.

One of the most important features of QCD dynamics though is chiral
symmetry breaking by a quark condensate, but since this 
is forbidden by unbroken
supersymmetry \footnote{A quark bilinear $\Psi\tilde\Psi$, where
$\Psi$ and $\tilde \Psi$ are fermionic components of chiral
superfields $Q = q + \theta \Psi \cdots, \tilde Q =\tilde q +
\theta \tilde \Psi + \cdots$, can be written as a SUSY variation
of another operator (it is an F-term of the composite operator 
$\tilde{Q}Q$).}, these constructions do not let us address
this issue. In the present paper we  attempt to come somewhat
closer to QCD by considering the embedding of D7-branes in two
non-supersymmetric backgrounds which exhibit confinement.
Although neither of these backgrounds corresponds exactly to QCD 
since they contain more degrees of freedom than just gluons and
quarks, we might nevertheless expect chiral symmetry breaking
behaviour. The quark mass $m$ and the quark condensate expectation value
$c$ are given by the
UV asymptotic behaviour of the solutions to the supergravity equations of
motion in the standard holographic way (see \cite{PolStrass} for an
example of this methodology). 
In the $\N=2$ supersymmetric scenario of
\cite{KarchKatz} with a D7 probe in standard AdS space, 
we show that there cannot be any regular
solution which has $c\neq 0$; the supersymmetric theory does not allow a quark
condensate. We then find that, for the deformed AdS 
backgrounds we consider, there are regular solutions with $c
\neq 0$. The case $c\neq 0$ with $m=0$ corresponds to spontaneous
chiral symmetry breaking.

The first supergravity
background we consider is the Schwarzschild black hole
in $AdS_5\times S^5$.  In the absence of 7-branes,  this
background is dual to strongly coupled ${\mathcal N} =4$ super
Yang-Mills at finite temperature and is in the same universality
class as three-dimensional pure QCD \cite{Witten}. Glueball
spectra in this case were computed in \cite{Csaki,Zyskin}. We introduce
D7-branes into this background and compute the quark condensate as
a function of the bare quark mass, as well as the meson spectrum.
This background is dual to the finite temperature version of the 
${\mathcal N} =
2$ Super Yang-Mills theory considered in
\cite{KarchKatz,MateosMyers}. The finite temperature
${\mathcal N} =2$ theory is not in the same universality class as
three-dimensional QCD with light quarks since the antiperiodic boundary
conditions for fermions in the Euclidean time direction give a
non-zero mass to the quarks upon reduction to three-dimensions,
even if the hypermultiplet mass of the underlying ${\mathcal N}=2$
theory vanishes.  In fact these quarks decouple if one takes the
temperature to infinity in order to obtain a truly three-dimensional theory.
Nevertheless at finite temperature the geometry describes an interesting
four-dimensional strongly coupled gauge configuration with quarks.
The meson spectrum we obtain has a mass gap of order of the
glueball mass. Furthermore we find that the  $\bar\Psi\Psi$
condensate vanishes for zero hypermultiplet mass, such that there is
no spontaneous
violation of parity in three dimensions or chiral symmetry in four dimensions. 
However for $m \neq 0$ we find a
condensate $c$  which at first grows linearly with $m$, and then
shrinks back towards zero. Increasing $m$ further, the D7
embedding undergoes a geometric transition at a critical mass
$m_c$. At sufficiently large $m \gg m_c$ the condenstate is
negligible and the spectrum matches smoothly with the one found in
\cite{MateosMyers} for the ${\mathcal N}=2$ theory. We speculate that
the geometric transition corresponds to a third order phase
transition in the dual gauge theory,  at which $\frac{dc}{dm}$
is discontinuous.

The second non-supersymmetric background which we  consider
was found by Constable and Myers \cite{CM}.  This
background is  asymptotically  $AdS_5 \times S^5$ but has a
non-constant dilaton and $S^5$ radius. In the field theory an
operator of dimension 4 with zero R-charge has been introduced.
This deformation does not give mass to the  adjoint fermions and 
scalars of the
underlying ${\mathcal N}=4$ theory but does leave a non-supersymmetric 
gauge background.
Furthermore,  unlike the AdS black-hole background,
the geometry has a naked singularity.
Nevertheless, in a certain parameter range, this background gives
an area law for the Wilson loop and a discrete spectrum of
glueballs with a mass gap. 

We obtain numerical solutions for the D7-brane equations of motion in
the Constable-Myers background
with asymptotic behaviour determined by a quark mass $m$ and
chiral condensate~$c$.   We compute the condensate $c$ as a function of
the quark mass $m$ subject to a regularity constraint. Remarkably,
our results are  not
sensitive to the singular behaviour of the metric in the IR.
For a given mass there are two regular solutions of which the physical, lowest
action solution corresponds to the D7 brane ``ending'' before reaching
the curvature singularity. Of course the D7-brane does
not really end, however the $S^3$ about which it is wrapped
contracts to zero size, similarly to the scenario discussed in
\cite{KarchKatz}.  In our case the screening of the singularity is
related to the existence of the condensate. Furthermore we find numerical
evidence for a non-zero condensate in the limit $m\rightarrow 0$.
This corresponds to spontaneous breaking of the $U(1)$
chiral symmetry which is non-anomalous in the large $N$ limit
\cite{Wittenetaprime} (for a review see \cite{Donoghue}).

We also compute the meson spectrum by studying classical
fluctuations about the D7-embedding.
For zero quark mass,  the meson spectrum
contains a massless mode, as it must due to the spontaneous
chiral symmetry breaking.
Note that since the spontaneously
broken axial symmetry is $U(1)$ for a single D7-brane,  the
associated Goldstone mode is a close cousin to the $\eta^{\prime}$
of QCD, which is a Goldstone boson in the large $N$ limit.
In principle, stringy corrections to supergravity should give the
$\eta^{\prime}$ a mass.  We  briefly comment on
generalizations to the case of more than one flavour or,
equivalently, more than one D7-brane. Moreover we give a holographic
version of the Goldstone theorem.

The main message of this paper is that non-supersymmetric gravity duals of 
gauge theories dynamically generate quark condensates and can break 
chiral symmetries. We stress that the physical interpretation of 
naked singularities is a
delicate issue, for instance in the light of the analysis of
\cite{Gubser:2000nd}. This applies in particular to the discussion of
light quarks and mesons. 
It is therefore an important part of 
our analysis that in the presence of a condensate the physical
solutions to the supergravity and DBI equations of motion never
reach the singularity in the IR. Of course it would be interesting to
understand this mechanism further and to see if it occurs in other
supergravity backgrounds as well.

The organisation of this paper is follows.  In section 2 we review
some of the previous results from the study of D7 probes in
the AdS/CFT Correspondence.  In section 3 we consider D7-branes in
the AdS Schwarzschild black hole background.  In section 4 we
study D7-brane probes in the Constable-Myers background
\cite{CM}.  In section 5 we conclude  and present
some open problems.

\section{AdS/CFT duality for an ${\mathcal N} =2$ gauge theory with 
fundamental matter}

It was first observed in \cite{KarchKatz} that one can obtain a
holographic dual of a four-dimensional Yang-Mills theory with
fundamental matter by taking a near-horizon
limit of a system of intersecting D3 and D7 branes. We  first
review some of the features of this duality, and then test
numerical techniques which we will use later to study deformations
of this duality.

Consider a stack of $N$ D3-branes spanning the directions $1,2,3$
and another stack of $N_f$ D7-branes spanning the directions
$1,2,3,4,5,6,7$. The low-energy dynamics of open strings in this
setting is described by a ${\mathcal N} =2$ super Yang-Mills
theory.  This theory contains the degrees of freedom of the
${\mathcal N} =4$ theory,  namely an ${\mathcal N} =2$ vector
multiplet and an ${\mathcal N}=2$ adjoint hypermultiplet,  as well
as $N_f$ ${\mathcal N} =2$ hypermultiplets in the fundamental
representation  of $SU(N)$. The theory is conformal in the limit
$N \rightarrow \infty$ with $N_f$ fixed.  There is an $SU(2)
\times U(1)$ R-symmetry.  The $U(1)$ R-symmetry
acts as a chiral rotation on the ``quarks'',  which are the
fermionic components of the ${\mathcal N} =2$ hypermultiplet
composed of fundamental and anti-fundamental chiral superfields
$Q$ and $\tilde Q$.  This symmetry also acts as a phase rotation
on the scalar component of one of the adjoint chiral superfields.
When the D7-branes are separated from the D3-branes in the two
mutually transverse directions $X^8$ and $X^9$,  the fields $Q,
\tilde Q$ become massive,  explicitly breaking the $U(1)$
R-symmetry and conformal invariance.  
%Flowing to the infrared then
%gives the ${\mathcal N} =4$ super Yang-Mills theory. 
As shown in
\cite{KarchKatz},  the ${\mathcal N} =2$ theory   as well as
its renormalization group flow have an elegant holographic
description.

This holographic description is obtained as follows.  In the limit
of large $N$ at fixed but large 't~Hooft coupling $\lambda = g^2N
>>1$, the D3-branes may be replaced with their near-horizon $AdS_5
\times S^5$ geometry. Since the number $N_f$ of D7-branes is
finite, their back-reaction on the geometry can be effectively
ignored. For determining the D7 probe brane embedding in $AdS_5 \times
S^5$, let us consider the D3-brane metric in the form
\begin{align}\label{metric}
ds^2 &= f(r)^{-1/2}(-dt^2 + d\vec x^2) + f(r)^{1/2}d\vec y^2 \, ,
\end{align} where $\vec x =
(X^1,X^2,X^3)$, $\vec y = (X^4, \cdots, X^9)$, $r^2 \equiv \vec
y^2$ and
\begin{align}
f(r) = 1 + \frac{R^4}{r^4} \, .
\end{align}
As usual, the $AdS_5 \times S^5$ geometry is obtained by dropping the $1$ in
$f(r)$ which is suitable in the near-horizon region $r/R <<1$.

For massless flavours, the D7 brane embedding in the D3-metric
(\ref{metric}) is given by $y^5 = y^6 =0$ (corresponding to
$X^8=X^9 =0$.).  In the $AdS_5 \times S^5$ geometry, the induced
metric on the D7-brane is $AdS_5 \times S^3$.  The D7-brane fills
$AdS_5$, while wrapping a great three-sphere of the $S^5$. The
isometries of the $AdS_5 \times S^5$ metric which preserve the
embedding correspond to the conformal group and R-symmetries of
the ${\mathcal N} =2$ gauge theory.   The conformal group
$SO(2,4)$ is the isometry group of $AdS_5$, while the
$SU(2) \times U(1)$ R-symmetry corresponds to the rotations of the
$S^3$ inside $S^5$ and rotations of the $y^5, y^6$ coordinates.

The holographic description for massive flavours is found by
considering the D7-brane embedding $y^5 = X^8 =0, y^6 = X^9 = m$.
In this case the D7-geometry is still $AdS_5 \times S^3$ in the
$r\rightarrow \infty$ region corresponding to the ultraviolet.
However as one decreases $r$,  the $S^3$ of the induced geometry
on the D7-brane contracts to zero size at $r=m$. This is possible
because the $S^3$ is contractible within the $S^5$ of the full ten
dimensional geometry. The D7-brane ``ends'' at the value of $r$ at
which the $S^3$ collapses,  meaning that it does not fill all of
$AdS_5$,  but only a region outside a core of radius $r=m$.  
%This
%is consistent with the fact that the theory contains no
%fundamental representations and is effectively ${\mathcal N} =4$
%super Yang-Mills at energies below $m$. 
Note that although the
D7-brane ends at $r=m$,  the D7-geometry is perfectly smooth, as
is illustrated in figure 1.  In the massive case,  the conformal
and $U(1)$ symmetries are broken,  and the D7 embedding
is no longer invariant under the corresponding
isometries.
\begin{figure}[!h]
\begin{center}
\includegraphics[height=6cm,clip=true,keepaspectratio=true]{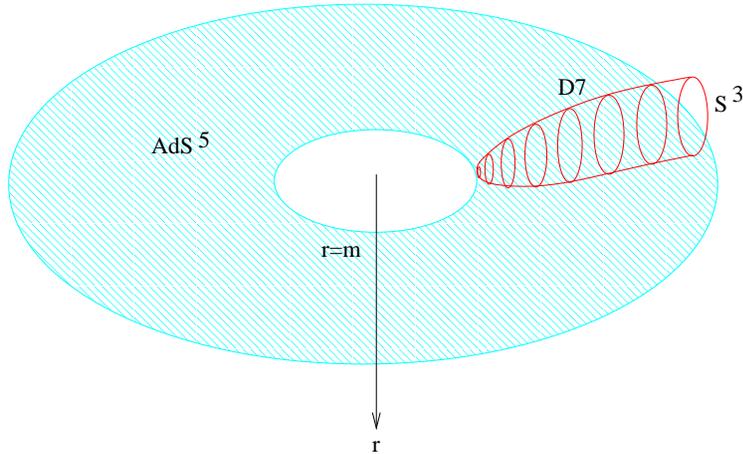}
\caption{The D7 embedding in $AdS_5 \times S^5$.}\label{flow}
\end{center}
\end{figure}

\subsection{Testing numerical methods}

In the subsequent sections we will numerically compute condensates
and meson spectra in deformations of the duality discussed above.
Therefore we  first test these numerical techniques against
some exact results in the undeformed case.

It will be convenient to write the transverse $d\vec y^2$ part of
the metric (\ref{metric}) in the following way \beq d\vec y^2 = d
\rho^2 +\rho^2 d \Omega_3^2 + dy_5^2 + dy_6^2, \eeq where
$d\Omega_3^2$ is a three-sphere metric.

To study the implications of the classical D7 probe dynamics for
the dual field theory, we now evaluate the scalar contributions to  the
Dirac-Born-Infeld (DBI) action for the D7 brane in the $AdS_5
\times S^5$ background. We work in static gauge where the world
volume coordinates of the brane are identified with the spacetime
coordinates by $\xi^a \sim t,x_1,x_2, x_3, y_1, ..., y_4$. The DBI
action is then
\begin{eqnarray}
S_{D7} & = & - T_7 \int d^8\xi \sqrt{- {\rm det}(P[G_{ab}] ) }
\nonumber \\ & = & - T_7 \int d^8 \xi ~ \epsilon_3 ~ \rho^3
\sqrt{1 + {\frac{g^{ab}}{  \rho^2 + y_5^2 + y_6^2}} (\partial_a y_5
\partial_b y_5 +
\partial_a y_6 \partial_b y_6)},
\end{eqnarray}
where $g_{ab}$ is the induced metric on the D7 brane and
$\epsilon_3$ is the determinant factor from the three sphere.
The ground state configuration of
the D7 brane is given by the solution of the Euler-Lagrange
equation with dependence only on the $\rho$ variable.
In this case the equations of motion become \beq \label{d7eom} {\frac{d}
{d \rho}} \left[ {\rho^3 \over  \sqrt{1 + \left({d y_6 \over d
\rho}\right)^2}}  { d y_6 \over d \rho}\right] = 0 \, , \eeq where we
consider solutions with $y_5=0$ only.  Recall
that the $U(1)$ R-symmetry corresponds to rotations in the
$y^5,y^6$ plane.

The equations of motion have asymptotic ($\rho\rightarrow \infty$)
solutions of the form \beq y_6 = m + { \frac{c}{ \rho^2}}. \eeq  
The identification of these constants as field theory operators
requires a coordinate transformation
because the scalar kinetic term is  not of the usual
canonical AdS form. Transforming to the coordinates of
\cite{KarchKatz} in which the kinetic term has canonical form, 
we see that $m$ has dimension 1
and $c$ has dimension 3. The scalars are then identified \cite{PolStrass}
with the
quark mass $m_q$ and condensate $\langle \bar{q} q \rangle$,
respectively, in agreement with the usual AdS/CFT dictionary.

Note that $y_6(\rho) = m$ is an exact solution of the equations of
motion,  corresponding to the embedding \cite{KarchKatz} reviewed above.
On the other hand there should be something ill-behaved about the
solutions with non-zero $c$,  since a quark condensate is
forbidden by supersymmetry.  In figure 2 we plot numerical solutions
of the equations of motion (obtained by a shooting technique using
Mathematica) for solutions with non-zero $c$, and find that they
are divergent.  The divergence of these solutions is not, by
itself, pathological because the variable $y_6$ is just the
location of the D7-brane.  However the AdS radius $r^2 = y_6(\rho)^2
+ \rho^2$ is not monotonically increasing as a function of $\rho$
for the divergent solutions. This means that these solutions have
no interpretation as a renormalization group flow,  or as a vacuum
of the dual field theory. As expected, the mass only solution is
the only well-behaved solution.
\begin{figure}[!h]
\begin{center}
\includegraphics[height=6cm,clip=true,keepaspectratio=true]{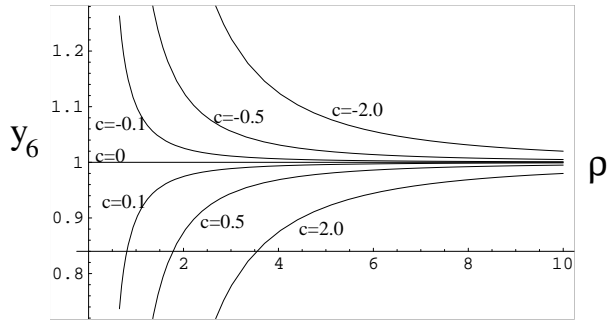}
\caption{- numerical solutions of EoM in AdS showing that in the
presence of a condensate asymptotically the solutions are divergent.
The regular solution is the mass only solution.}\label{numsol1}
\end{center}
\end{figure}

The other exact result which we wish to test numerically is the
meson spectrum. In order to find the states with zero spin on the
$S^3$, one looks for normalizable solutions of the equations of
motion of the form \beq y_6 + i y_5 = m + \delta  (\rho) e^{i k
x}, \hspace{1cm} M^2 = - k^2 \eeq where one linearizes in the
small fluctuation $\delta(\rho)$. The linearized equation of
motion is \beq
\partial_\rho^2 \delta(\rho)  + { \frac{3}{  \rho}} \partial_\rho \delta(\rho)  
+
{\frac{M^2}{  (\rho^2+m^2)^2}} \delta(\rho)  = 0. \eeq This was solved
exactly in~\cite{MateosMyers}, where it was shown that the
solutions can be written as hypergeometric functions \beq
\delta(\rho) = {\frac{A}{   (\rho^2 + 1)^{n+1}}} F(-n-1,-n;2,-\rho^2)
\eeq with $A$ a constant. The exact mass spectrum is then given by
\beq \label{formula} M = 2 m \sqrt{(n+1)(n+2)}, \hspace{1cm}
n=0,1,2, \dots \, . \eeq

We are interested in whether we can reproduce this result
numerically, via a shooting technique. The equation of motion can
be solved numerically subject to boundary conditions $\delta(\rho)
\sim c/\rho^2$ at large $\rho$ indicating that the meson is a
quark bilinear of dimension 3 in the UV. Solutions of the equation
must be regular at all $\rho$ so the allowed $M^2$ solutions can
be found by tuning to these regular forms. The result
(\ref{formula}) is easily reproduced to 2 significant figures. We
show an example of the method in action in figure 3.
\begin{figure}[!h]
\begin{center}
\includegraphics[height=6cm,clip=true,keepaspectratio=true]{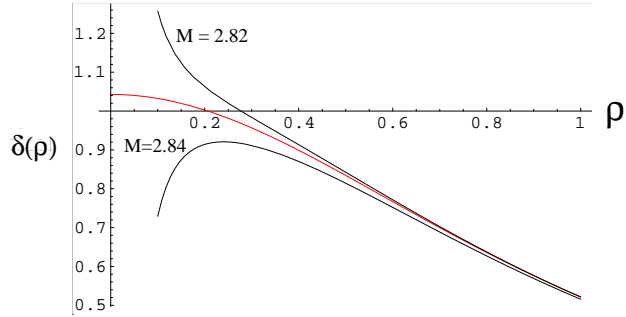}
\caption{- numerical solutions of the meson EoM for different values of
$M$ showing the identification
of the first bound state mass - the exact regular solution is plotted 
between
the two numerical flows.}\label{numsol}
\end{center}
\end{figure}

\section{The AdS-Schwarzschild Solution}

\subsection{The background}

We now move on to study quark condensates and mesons in a
non-supersymmetric deformation of the AdS/CFT correspondence and
study the AdS-Schwarzschild black hole solution. This
geometry is dual to the ${\cal N}=4$ gauge theory at finite
temperature \cite{Witten},  which is in the same universality
class as pure three-dimensional QCD.

The Euclidean AdS-Schwarzschild solution is given by \beq
\label{bh}ds^2 = K(r) d\tau^2 + {\frac{dr^2}{ K(r)}} + r^2
dx_{\parallel}^2 + d\Omega_5^2 \, , \eeq where \beq K(r) = r^2 - {\frac{b^4}
{r^2}} \,. \eeq This space-time is smooth and complete if
$\tau$ is periodic with period $\pi b$. Note that the $S^1$
parametrized by $\tau$ collapses at $r= b$.  The fact that the
geometry ``ends'' at $r=b$ is responsible for the existence of an
area law for the Wilson loop and a mass gap in the dual field
theory (see \cite{Witten}).  The period of $\tau$ is
equivalent to the temperature in the dual ${\mathcal N} =4$ gauge
theory. The parameter $b$ sets the scale of the deformation and
for convenience in the numerical work below we shall set it equal
to 1.  At finite temperature, the fermions have anti-periodic
boundary conditions in the Euclidean time direction and become
massive upon dimensional reduction to three dimensions. The
adjoint scalars also become massive at one loop. Thus in the 
high-temperature limit,  the adjoint fermions and scalars decouple,
leaving pure three-dimensional QCD.

We  now introduce a D7-brane into this background,  which
corresponds to the addition of matter in the fundamental
representation.  The dual gauge theory is the ${\mathcal N}=2$
gauge theory of Karch and Katz at finite temperature.  Note that
the fermions in the fundamental representation also have
anti-periodic boundary conditions in the Euclidean time direction.
Thus these also decouple in the high-temperature limit,  as do the
fundamental scalars which get masses at one loop,  leaving pure
QCD$_3$ as before.  Thus in this particular case we are not
interested in the high-temperature limit, but only the region
accessible to supergravity and Dirac-Born-Infeld theory. Although
the dual field theory cannot be viewed as a three-dimensional
gauge theory with light quarks, it is nevertheless a four
dimensional non-supersymmetric gauge theory with confined degrees
of freedom in the fundamental representation. This provides an
interesting, if exotic, setting to compute quark condensates and
meson spectra using Dirac-Born-Infeld theory.  The Constable-Myers
background which we will consider later turns out to have more
realistic properties.

\subsection{Embedding of a D7 brane}

To embed a D7 brane in the AdS black-hole background it is
useful to recast the metric (\ref{bh}) to a form with an explicit
flat 6-plane. To this end, we change variables from $r$ to $w$,
such that
\begin{equation}
{\frac{dw}{ w}}\equiv {\frac{r dr}{ (r^4-b^4)^{1/2}}} \, , 
\end{equation}
which is  solved by
\begin{equation}
2 w^2=r^2+\sqrt{r^4-b^4} \,.
\end{equation}
The metric is then
\begin{equation}
ds^2=\left(w^2+\frac{b^4}{4w^2}\right)d\vec
x^2+\frac{(4w^4-b^4)^2}{4w^2(4w^4 +b^4)}dt^2+
\frac{1}{w^2}(\sum_{i=1}^6 dw_i^2) \, , 
\end{equation}
where $\sum_i dw_i^2 = dw^2 + w^2 d\Omega_5^2$,  which for reasons
of convenience will also be written as $d\rho^2 +
\rho^2d\Omega_3^2 + dw_5^2 + dw_6^2$ where $d\Omega_3^2$ is the
unit three-sphere metric.
%It is now
%straightforward to embed a spacetime filling D7 brane into the
%AdS-Schwarzschild background.
The AdS black hole geometry asymptotically approaches $AdS_5
\times S^5$ at large $w$.  Here the background becomes
supersymmetric,  and the D7 embedding should approach that
discussed in \cite{KarchKatz}.  The asymptotic solution has the
form $w_6 = m, w_5 = 0$,  where $m$ should be interpreted as a
bare quark mass. To take into account the deformation,  we will
consider a more general ansatz for the embedding of the form $w_6
= w_6(\rho), w_5=0$,  with the function $w_6(\rho)$ to be
determined numerically.
%Stability of the embedding is guaranteed by
%the asymptotic AdS behaviour of the background.
The DBI action for the orthogonal directions $w_5, w_6$ is \beq
S_{D7} =-\mu_7\int d^8\xi ~\epsilon_3 ~ {\cal G}(\rho,w_5,w_6)
\left( 1 + { g^{ab} \over (\rho^2 + w_5^2 + w_6^2)} \partial_a w_5
\partial_b
w_5 + { g^{ab} \over (\rho^2 + w_5^2 + w_6^2)} \partial_a w_6 \partial_b
w_6 \right)^{1/2}
\eeq
where the determinant of the metric is given by

\beq
{\cal G}(\rho,w_5,w_6) = \sqrt{ g_{tt} g_{xx}^3 \rho^6 \over (\rho^2 + w_5^2
+ w_6^2)^4}
= \rho^3 { ( 4(\rho^2 + w_5^2 + w_6^2)^2 + b^4) ( 4( \rho^2
+ w_5^2 + w_6^2)^2 - b^4) \over 16 (\rho^2 + w_5^2 + w_6^2)^4} \,.
\eeq

With the ansatz $w_5=0$ and $w_6 = w_6(\rho)$, the equation of
motion becomes

\beq\label{eqnmot} {d \over d\rho} \left[ {\cal G}(\rho,w_6)
\sqrt{ 1 \over 1 + \left({  d w_6 \over d \rho}\right)^2} { d w_6
\over d \rho} \right] - \sqrt{1 + \left({  d w_6 \over d
\rho}\right)^2} {b^8 \rho^3 w_6 \over 2 ( \rho^2 + w_6^2)^5} = 0
\,. \eeq
%The supersymmetry breaking manifests as a potential for $w_6$.
The solutions
of this equation determine the induced metric on the D7 brane which
is given by

\begin{equation} ds^2 = \left( \tilde w^2 +
  \frac{b^4}{4 \tilde w^2} \right) d\vec x^2 + \frac{(4 \tilde w^4 -
  b^4)^2}{4\tilde w^2 (4\tilde w^4 +b^4)} dt^2 + \frac{1+(\partial_\rho
w_6)^2}{\tilde w^2} d
\rho^2
+ \frac{\rho^2}{\tilde w^2} d\Omega_3^2 \, , \label{D7metric}
\end{equation}
with $\tilde w^2=\rho^2+w_6^2(\rho)$.
%Since the black hole geometry is
%$AdS_5
%\times S^5$ asymptotically, the solutions approach the constant solutions
%\cite{MateosMyers} $w_6=m$ at $\rho \rightarrow \infty$ and
The D7 brane metric becomes $AdS_5 \times S^3$ for $\rho \gg b,m$.

\subsection{Karch-Katz solutions vs condensate solutions}

Before computing the explicit D7-brane solutions, we remark that
there are several possibilities for the topology of the D7-brane
embedding which we illustrate in Fig 4.

\begin{figure}[!h]
\begin{center}
\includegraphics[height=7cm,clip=true,keepaspectratio=true]{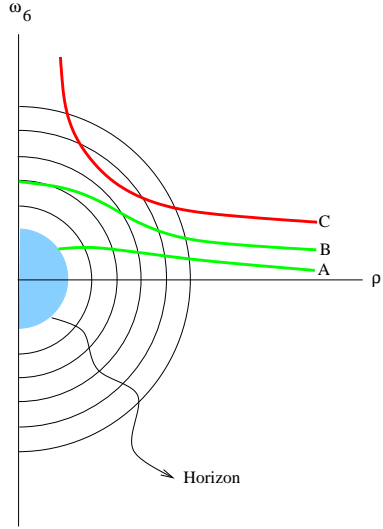}
\caption{\small{Different possibilities for solutions of the
D7-brane equations of motion.  The semicircles are lines of
constant $r$, which should be interpreted as a scale in the dual
Yang-Mills theory. The curves of type $A,B$ have an
interpretations as an RG flow, while the curve $C$ does not.}
}\label{quiv}
\end{center}
\end{figure}

The UV asymptotic (large $\rho$) solution, where the geometry returns
to $AdS_5\times S^5$,  is of the form \beq
\label{asymp} w_6(\rho) \sim m + \frac{c}{\rho^2}.\eeq The
parameters $m$ and $c$ have the interpretation as a quark mass and
bilinear quark condensate respectively, as discussed below equation
(6).  These parameters can be
taken as the boundary conditions for the second order differential
equation (\ref{eqnmot}),  which we solve using a numerical
shooting technique.  Of course the physical solutions should not
have arbitrary $m$ and $c$. The condition which we  use to
identify physical solutions is that the D7-brane embedding should
have an interpretation as a RG flow.  This implies for instance
that if one slices the D7-brane geometry at a fixed value of
$r^2$,  or equivalently at a fixed value of $w_6^2 + \rho^2$, one
should obtain at most one copy of the geometry $R^4 \times S^3$.
In other words,  $r^2 = \rho^2 + w_6(\rho)^2$ should be a
monotonically increasing function of $\rho$.  This is certainly
not the case for divergent solutions.   Such solutions are not
in correspondence with a vacuum of the dual gauge theory and are
discarded.

There are then two possible forms of regular solutions.
The geometry in which the D7-brane is embedded has the
boundary topology $R^3 \times S^1 \times S^5$, which contains the
D7-brane boundary $R^3 \times S^1 \times S^3$. Recall that the
$S^3$ is contractible within $S^5$. Furthermore the $S^1$ is
contractible within the bulk geometry and shrinks to zero as one
approaches the horizon $r=b$.  The D7-brane may either ``end'' at
some $r > b$ if the $S^3$ collapses, or it may continue all the
way to the horizon where the $S^1$ collapses but the $S^3$ has
finite size. In other words, the D7-topology may be either $R^3
\times B^4 \times S^1$ or $R^3 \times S^3 \times B^2$. The former
is the type found in \cite{KarchKatz}.  As one might expect,  this
topology occurs when the quark mass $m$ is sufficiently large
compared to $b$. In this case, the $S^3$ of the D7-brane contracts
to zero size in the asymptotic region where the deformation of
$AdS_5 \times S^5$ is negligible. We shall find the other topology
for sufficiently small $m$.

For any chosen value of $m$,  we find only a discrete choice of
$c$ which gives a regular solution that can be interpreted as an
RG flow. In Figure 5 we show sample numerical flows used to identify a 
regular solution. 
\begin{figure}[!h]
\begin{center}
\includegraphics[height=7cm,clip=true,keepaspectratio=true]{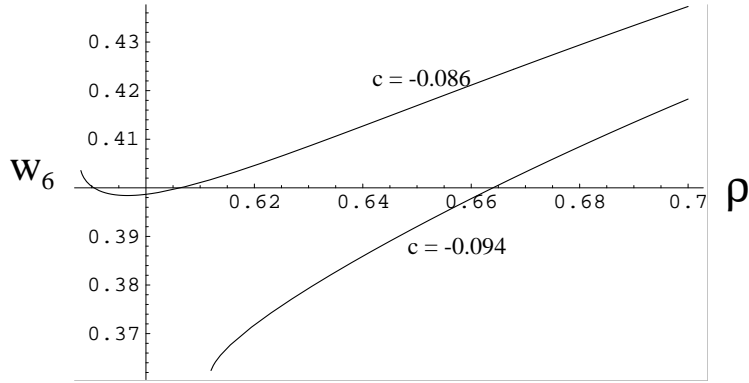}
\caption{An example (for
$m=0.6$) of the different flow behaviours around the regular
(physical) solution.}\label{quiv4}
\end{center}
\end{figure}

For the regular  solutions the D7-brane either ends at the
horizon, \beq w_6^2 + \rho^2 = \frac{1}{2}b^2,\eeq at which the
$S^1$ collapses, or ends at a point outside the horizon, \beq \rho
=0, w_6^2 \ge \frac{1}{2}b^2,\eeq at which the $S^3$ collapses
(see (\ref{D7metric})).  Both types of solution are illustrated in
figure 6
for several choices of $m$.  
We choose units such that $b=1$. We
refer to solutions with collapsing $S^3$ as Karch-Katz solutions,
and solutions with collapsing $S^1$ as ``condensate'' solutions,
for reasons that will become apparent shortly.
\begin{figure}[!h]
\begin{center}
\includegraphics[height=7cm,clip=true,keepaspectratio=true]{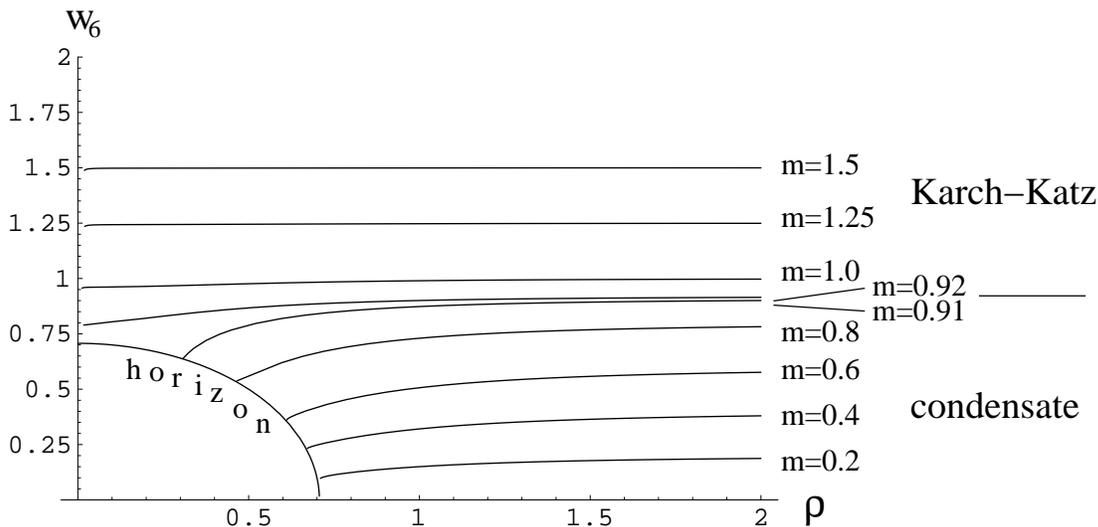}
\caption{Two classes of regular solutions in the AdS black hole
background.}\label{quiv2}
\end{center}
\end{figure}
Note that the boundary between the Karch-Katz and condensate
solutions is  at a critical value of the mass $m=m_{\rm crit}$
such that $w_6(\rho=0)=\sqrt{1/2} b$.  In this case both the $S^1$
and $S^3$ collapse simultaneously. Numerically, we find $m_{\rm
crit} \approx 0.92$.

There is an exact solution of the equation of motion $w_6 = 0$
which is regular and corresponds to $m=c=0$. Thus there is no
condensate when the quarks are massless (from the four-dimensional
point of view). This should not be disappointing,  since the
theory is not in the same universality class as QCD$_3$ with light
quarks. The quarks obtain a mass of order the temperature which will
tend to suppress the formation of a condensate. 
For non-zero $m$ we obtain the solutions numerically. The
dependence of the condensate on the mass is illustrated in figure
7. We find that as $m$ increases, the condensate $c$ initially
increases and then decreases again. At sufficiently large $m$, the
condensate becomes negligible,  which is to be expected as the
D7-brane ends in the region where the deformation of AdS is small.
Recall that there is no condensate in the Yang-Mills theory with
unbroken ${\mathcal N} =2$ supersymmetry described by D7-branes in
un-deformed AdS.

\begin{figure}[!h]
\begin{center}
\includegraphics[height=7cm,clip=true,keepaspectratio=true]{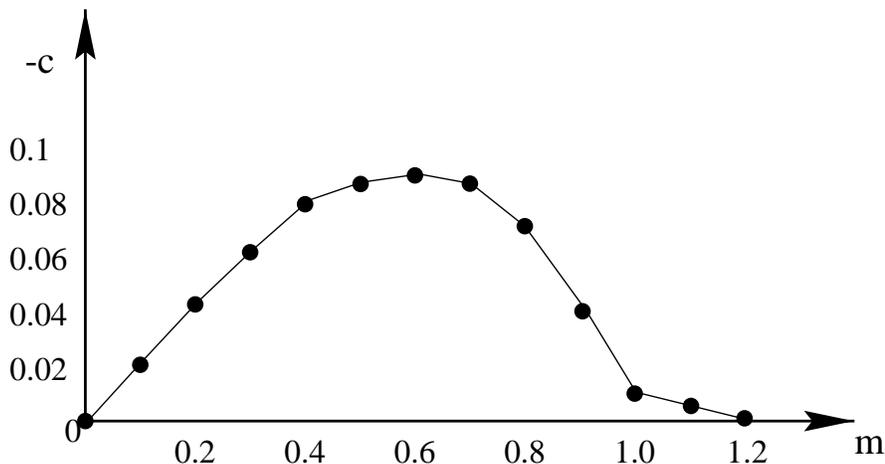}
\caption{A plot of the parameter $c$ vs $m$ for the regular
solutions in AdS Schwarzschild. The linear fit between points is
just to guide the eye.}\label{quiv3}
\end{center}
\end{figure}

Since the D7-brane topology changes as one crosses $m_{\rm crit}$,
one might expect a phase transition to occur at this
point\footnote{For $m>m_{\rm crit}$ there is the interesting
possibility of introducing an even spin structure on the $S^1$ of
the D7-brane,  since this $S^1$ is no longer contractible on the
D7.  If this is a sensible (i.e stable) background,  it would
correspond to a different field theory, in which the fundamental
fermions are periodic on $S^1$ and do not have a Kaluza-Klein
mass. We have not analyzed this possibility.}. Looking at figure 7, 
there does not appear to be any discontinuity in $c(m)$ 
at $m = m_{\rm crit}$. It
is still possible that there is a third order transition at this
point, corresponding to a discontinuity in the slope $\frac{dc}{dm}$.
Of course the numerical results must be refined significantly to
find evidence for such a transition.

\subsection{Meson spectrum}

The meson spectrum can be found by solving the linearized
equations of motion for small fluctuations about the D7-embeddings
found above.  Let us consider the fluctuations of the variable
$w_5$ about the embedding,  which has $w_5=0$.  We take \beq w_5 =
f(\rho)\sin( k \cdot \vec x). \eeq The linearized (in $w_5$)
equation of motion is \beq \label{meep} \begin{array}{c} {d \over
d\rho} \left[ {\cal G}(\rho, w_6) \sqrt{ 1 \over 1 + \left({  d
w_6 \over d \rho}\right)^2} { d f(\rho) \over d \rho} \right] +
{\cal G}(\rho, w_6) \sqrt{ 1 \over 1 + \left({  d w_6 \over d
\rho}\right)^2} \left({ 4 \over 4 (\rho^2 +w_6^2)^2 + b^4}\right)
M^2 f(\rho)
\\
\\
-  \sqrt{1 + \left({  d w_6 \over d \rho}\right)^2} { b^8 \rho^3
f(\rho) \over 2 ( \rho^2 + w_6^2)^5} = 0 \,.
\end{array}
\eeq where $M^2 = \vec k^2$. The allowed values of $\vec k^2$ are
determined by requiring the solution to be normalizable and
regular. Note that if the $U(1)$ symmetry which rotates $w_5$ and
$w_6$ were spontaneously broken by an embedding of the asymptotic
form $w_6 \sim c/\rho^2$ with non-zero $c$,  there would be a
massless state in the spectrum associated with $w_5$ fluctuations.
This is not the case  in the present setting, since the condensate
is only non-zero for non-zero quark mass $m$. Instead we find a
mass gap in the meson spectrum.  We have computed the meson
spectrum by solving (\ref{meep}) by a numerical shooting techique.
As in the Karch-Katz geometry, we seek regular solutions for $w_5$
which are asymptotically of the form $c/\rho^2$ in the presence of
the background $w_6$ solution. The results for the meson masses
are plotted in figure 8. Of course the meson mass gap here can
be largely attributed to the Kaluza-Klein masses of the
constituent quarks, which are of the same order as the temperature
($T \sim \pi$ in units with $b=1$).

Thus we have seen that, while the thermal gauge background allows
a quark condensate when it does not spontaneously break any
symmetries, there is no chiral (parity) symmetry breaking at zero quark
mass. The meson spectrum reflects this by having a mass gap - the
fermions have an induced mass from the presence of finite
temperature. In the subsequent discussion we will consider another
background which admits light constituent quarks and has
properties much closer to QCD.

\begin{figure}[!h]
\begin{center}
\includegraphics[height=7cm,clip=true,keepaspectratio=true]{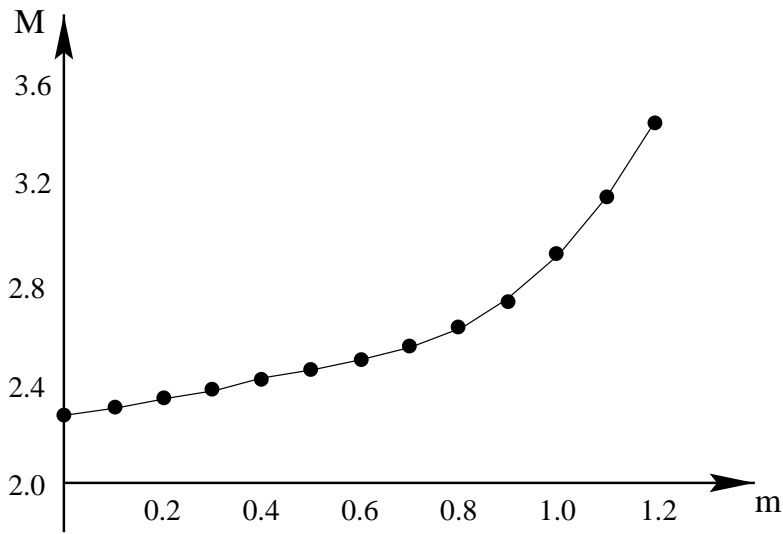}
\caption{A plot of the $w_5$ meson mass vs $m$ in AdS
Schwarzschild. The linear fit between points is just to guide the eye.}\label{quiv5}
\end{center}
\end{figure}

%%%%%%%%%%%%%%%%%%%%%%%%%%%

\section{The Constable-Myers Deformation}

\subsection{The background}

We consider the non-supersymmetric deformed AdS geometry
originally constructed in \cite{CM}. This geometry
corresponds to the ${\mathcal N}=4$ super Yang-Mills theory
deformed by the presence of a vacuum expectation value for an
R-singlet operator with dimension four (such as $tr F^{\mu
\nu}F_{\mu \nu}$). The supergravity background has a dilaton and
$S^5$ volume factor depending on the radial direction. In a
certain parameter range, this background implies an area law for
the Wilson loop and a mass-gap in the glueball spectrum. Whether
the geometry,  which has a naked singularity, actually describes
the stable non-supersymmetric vacuum of a field theory is not well
understood \cite{CM}.  This is not so important from our
point of view though since the geometry is a well defined gravity
description of a non-supersymmetric gauge configuration. We can
just ask about the behaviour of quarks in that background.

The geometry in {\it Einstein frame} is given by

\beq ds^2 = H^{-1/2} \left( 1 + { 2 b^4 \over r^4}
\right)^{\delta/4} dx_{4}^2 + H^{1/2} \left( 1 + { 2 b^4 \over
r^4} \right)^{(2-\delta)/4} {r^2 \over \left( 1 + {b^4 \over
r^4}\right)^{1/2}} \left[ {r^6 \over (r^4 + b^4)^2} dr^2 + d
\Omega^2_5 \right], \eeq where

\beq H =  \left( 1 + { 2 b^4 \over
r^4} \right)^{\delta} - 1
\eeq
and with the dilaton and four-form given by

\beq
e^{2 \phi} = e^{2 \phi_0} \left( 1 + { 2 b^4 \over r^4}
\right)^{\Delta}, \hspace{1cm} C_{(4)} = - {1 \over 4} H^{-1} dt
\wedge dx \wedge dy \wedge dz \,.
\eeq
The parameter $b$ corresponds
to the vev of the dimension 4 operator. The parameters $\Delta$
and  $\delta$ are constrained by

\beq \Delta^2 + \delta^2 = 10. \eeq Asymptotically the AdS
curvature is given by $L^4 = 2 \delta b^4$, so it makes sense to
set (with $L=1$)

\beq \delta = {1 \over 2 b^4}. \eeq $b$ is the only free parameter
in the geometry and its value sets the scale of the conformal
symmetry and supersymmetry breaking. We will again numerically set
it equal to 1 below.

To embed a D7 brane in this background it will again be convenient
to recast the metric in a form containing an explicit flat
6-plane. To this end, we change variables from $r$ to $w$, such that

\begin{equation}
\frac{dw}{w}\equiv \frac{r^2d(r^2)}{2(r^4+b^4)},
\end{equation}
which is solved by

\begin{equation}
\ln (w/w_0)^4 =\ln (r^4+b^4)
\end{equation}
or
\begin{equation}
(w/w_0)^4=r^4+b^4.
\end{equation}
So for the case of $b=0$ we should set the integration constant
$w_0=1$. The full metric is now

\beq ds^2 = H^{-1/2} \left( { w^4 + b^4 \over w^4-b^4}
\right)^{\delta/4} dx_{4}^2 + H^{1/2} \left( {w^4 + b^4 \over w^4-
b^4}\right)^{(2-\delta)/4} {w^4 - b^4 \over w^4 } \sum_{i=1}^6
dw_i^2, \eeq where

\beq H =  \left(  { w^4
+ b^4 \over w^4 - b^4}\right)^{\delta} - 1
\eeq
and the dilaton and four-form become

\beq e^{2 \phi} = e^{2 \phi_0} \left( { w^4 + b^4 \over
w^4 - b^4} \right)^{\Delta}, \hspace{1cm} C_{(4)} = - {1 \over 4}
H^{-1} dt \wedge dx \wedge dy \wedge dz.
\eeq

We now consider the D7 brane-action in the static gauge with
world-volume coordinates identified with the four Minkowski
coordinates - denoted by $x_{4}$ - and with
$w_{1,2,3,4}$. The transverse fluctuations are parameterized by
$w_5$ and $w_6$.  It is again convenient to define a coordinate
$\rho$ such that $\sum_{i=1}^4 dw_i^2 = d\rho^2 + \rho^2
d\Omega_3^2$. The DBI action in Einstein frame

\beq S_{D7} = - T_7 \int d^8 \xi e^{\phi} \sqrt{- {\rm
det}(P[G_{ab}])} \eeq can then be written as \beq S_{D7} = - T_7
\int d^8 \xi~ \epsilon_3 ~  e^{\phi} { \cal G}(\rho,w_5,w_6)
\left( 1 + g^{ab} g_{55} \partial_a w_5
\partial_b w_5 + g^{ab} g_{66} \partial_a w_6
\partial_b w_6 \right)^{1/2},
\eeq
where

\beq {\cal G}(\rho,w_5,w_6) = \rho^3 {( (\rho^2 + w_5^2
+ w_6^2)^2 + b^4) ( (\rho^2 + w_5^2 + w_6^2)^2 - b^4) \over
(\rho^2 + w_5^2 + w_6^2)^4}.
\eeq
We  again look for classical solutions to the EoM of the form
$w_6 = w_6(\rho), w_5 =0$, that define the ground state. They satisfy
\beq \label{eommc}{ d \over d \rho} \left[ {e^{\phi}  { \cal
G}(\rho,w_6) \over \sqrt{ 1 + (\partial_\rho w_6)^2}}
(\partial_\rho w_6)\right] - \sqrt{ 1 + (\partial_\rho w_6)^2} { d
\over d w_6} \left[ e^{\phi} { \cal G}(\rho,w_6) \right] = 0\,.
\eeq The last term in the above is a ``potential" like term that
is evaluated to be

\beq { d \over d w_6} \left[ e^{\phi} { \cal G}(\rho,w_6)
\right] = { 4 b^4 \rho^3 w_6 \over (\rho^2 + w_6^2)^5} \left( {
(\rho^2 + w_6^2)^2 + b^4 \over  (\rho^2 + w_6^2)^2 - b^4}
\right)^{\Delta/2} (2 b^4  - \Delta (\rho^2 + w_6^2)^2)\,. \eeq\\

We now consider numerical solutions with the asymptotic behavior
$w_6 \sim m + c/\rho^2$, and find the physical solutions by
imposing a regularity constraint as discussed in the previous
section. Note that unlike the Euclidean AdS black hole,  the
Constable-Myers background has a naked singularity at $r=0$ or
$\rho^2 + w_6^2 = b^2$. Thus there are  two possibilities for a
solution with an interpretation as an RG flow, which are as
follows. Either the D7-brane terminates at a value of $r$ away
from the naked singularity via a collapse of the $S^3$, or the
D7-brane goes all the way to the singularity.  In the latter case
we would have little control over the physics without a better
understanding of string theory in such highly curved backgrounds.
Different possibilities for solutions of the D7-brane equations of
motion are illustrated in figure 9.

\begin{figure}[!h]
\begin{center}
\includegraphics[height=7cm,clip=true,keepaspectratio=true]{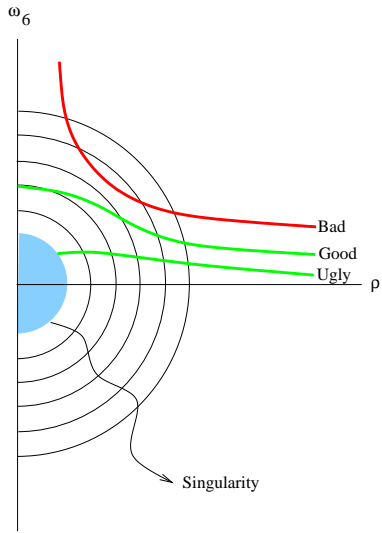}
\caption{\small{Different possibilities for solutions of the
D7-brane equations of motion.  The semicircles are lines of
constant $r$, which should be interpreted as a scale in the dual
Yang-Mills theory. The ``Bad'' curve cannot be interpreted as an
RG flow.  The other curves have an RG flow interpretation, however
the infrared (small r) region of the ``Ugly'' curve is outside the
range of validity of DBI/supergravity.}
}\label{quiv6}
\end{center}
\end{figure}

Fortunately something remarkable happens.  For positive
values of $m$ we find that there is a discreet regular solution
for each value of the mass that terminates at $w\geq 1.3$
before reaching the singularity at $w=1$.  Some of these regular solutions are
plotted in figure 10.  For $m=0$,  the solution $w_6 =0$ is exact,
which naively seems to indicate the absence of a chiral condensate
$(c=0)$. However, this solution reaches the singularity, and
therefore cannot be trusted.  On the other hand for a very small
but non-zero mass, the regular solutions require a non-vanishing
$c$ and terminate before reaching the singularity! The numerical
evidence (see figure 10) suggests that there is a non-zero
condensate in the limit $m\rightarrow 0$. In other words the
geometry spontaneously breaks the U(1) chiral symmetry. We plot
$c$ as a function of $m$ in figure 11.  This seems to be analogous
to the situation in field theory,  in which the path integral {\it
formally} gives no spontaneous symmetry breaking, which is found
only in the limit that a small explicit symmetry breaking
parameter is taken to zero.

\begin{figure}[!h]
\begin{center}
\includegraphics[height=7cm,clip=true,keepaspectratio=true]{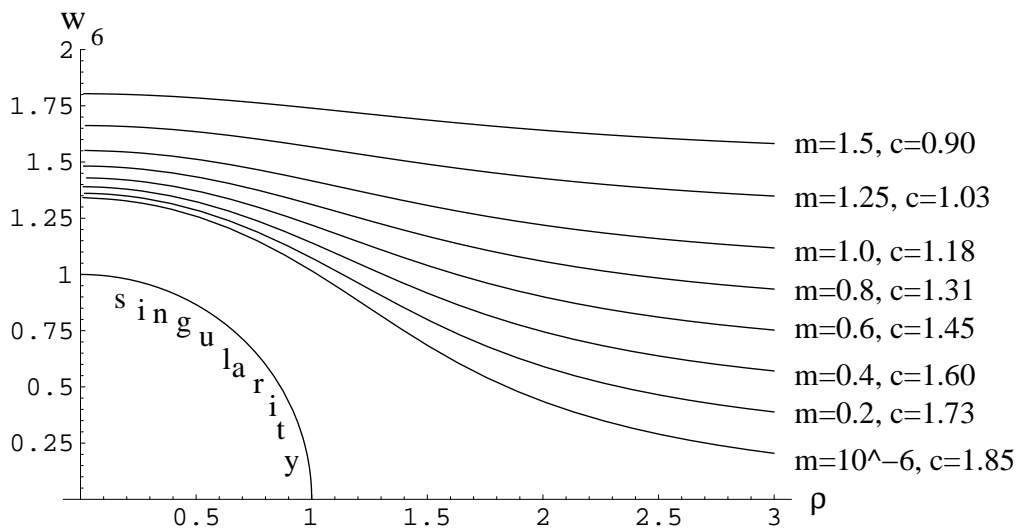}
\caption{Regular solutions in the Constable-Myers background.}
\label{quiv7}
\end{center}
\end{figure}

\begin{figure}[!h]
\begin{center}
\includegraphics[height=7cm,clip=true,keepaspectratio=true]{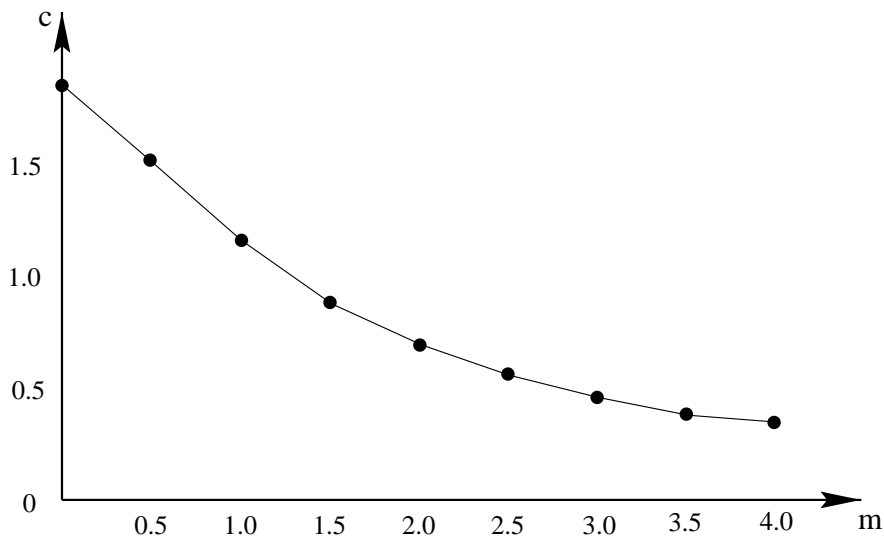}
\caption{A plot of the condensate parameter $c$ vs quark mass $m$ for
the regular solutions of the equation of motion in the Constable-Myers
background.}
\label{quiv8}
\end{center}
\end{figure}

We can study this phenomenon further in the deep infrared, in
particular in view of gaining further understanding of the behaviour
of the solutions shown in figure 10. Consider 
(\ref{eommc}) as $\rho \rightarrow 0$ with $w_6 \neq 0$ - the dilaton
becomes $\rho$ independent whilst ${\cal G} \sim \rho^3$. Thus the
potential term vanishes as $\rho^3$ whilst the derivative piece contains
a term that behaves like $\rho^2 \partial_\rho w_6$ and dominates.
Clearly there is a solution where $w_6$ is just a constant. This is the
regular behaviour we are numerically tuning to. It is now easy to find
the flows in reverse by setting the infrared constant value of $w_6$ and 
numerically solving out to the UV. This method ensures that the flow is 
always regular. We have checked that the asymptotic values of
the condensate as a function of mass match our previous computation at 
the level of a percent, showing that the numerics are under control.

\begin{figure}[!h]
\begin{center}
\includegraphics[height=7cm,clip=true,keepaspectratio=true]{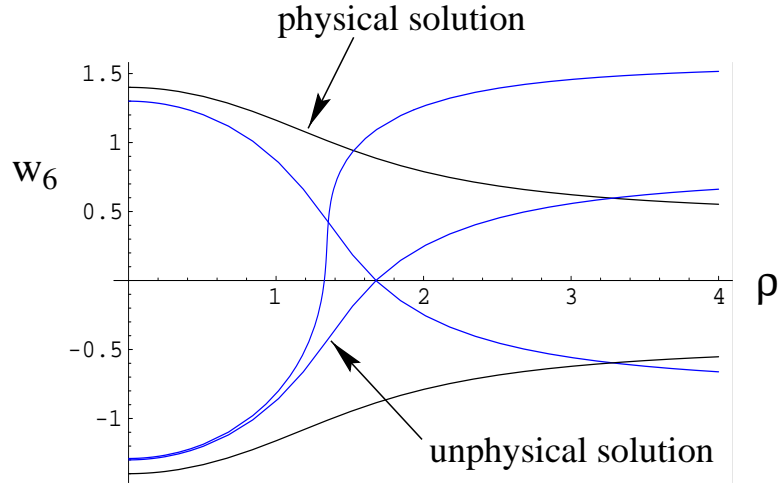}
\caption{Regular D7 embedding solutions in the Constable-Myers geometry
which lie close to the singularity in the infra-red.}
\label{new1}
\end{center}
\end{figure}

\begin{figure}[!h]
\begin{center}
\includegraphics[height=7cm,clip=true,keepaspectratio=true]{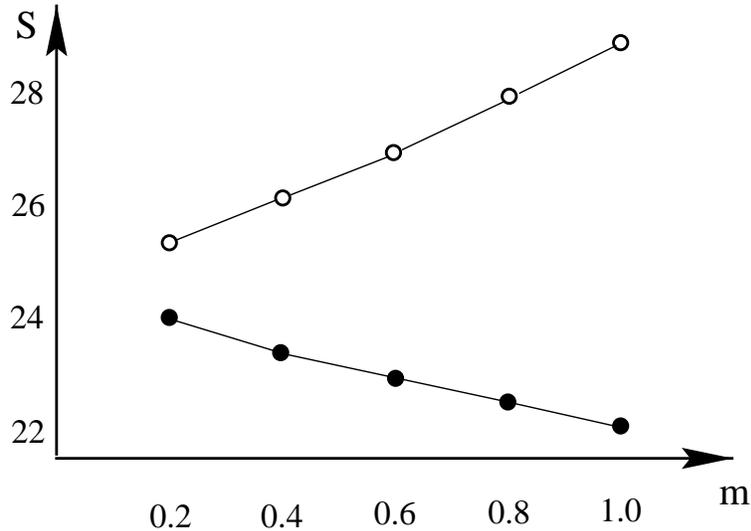}
\caption{A plot of action vs mass for the two regular D7 embedding
solutions in the Constable-Myers geometry. The higher action solutions
correspond to the flows that end at $|w_6| < 1.3$.}
\label{new2}
\end{center}
\end{figure}

We now realize though that there are infrared solutions where
 $w_6(\rho=0) < 1.3$.
These flows lie close to the singularity at $w_6(\rho=0) = 1 $ which we 
had hoped to
exclude. Solving these flows numerically we find that they 
flow to negative masses in the ultraviolet. We show these flows in
figure 12. Since there is a $w_6 \rightarrow - w_6$ symmetry of the solution
though this means there is a second regular flow for each positive mass
which in the infra-red flows to negative values, also shown in figure 12.
The flows that begin closer to the singularity in the infrared flow out
to larger masses in the ultraviolet. This strongly suggests these 
flows are not physical. When the quarks have a large mass, relative to the 
scale of the deformation, we do not expect the infra-red dynamics to have
a large influence on the physics. Thus the flows shown in figure 10 that
match onto the Karch-Katz type solution for large mass are the expected
physical solutions. To find some analytic support for this conjecture
we have calculated the action of two of the solutions for each mass value.
The action is formally infinite if we let the flow cover the whole space. 
To see the difference in action we have calculated the contribution
for $0< \rho<3$ only which covers the infrared part of the solution
(varying the upper limit does not change the conclusions). We plot the action
of the two solutions  versus the quark mass in figure 13,
from which it can be seen that the action for the solution
lying closer to the singularity is larger. Therefore the corresponding
solution is not relevant for the physics.

In a certain sense, the condensate screens the probe physics from
the naked singularity.
In the limit of small explicit symmetry breaking parameter $m$,
any solution (or vacuum) which did not break the $U(1)$ symmetry
would have $w_6\rightarrow 0$ for all $\rho$.  If this were the
case, the solution would reach the singularity (see figure 9).
The screening of the singularity is reminiscent of the
enhan\c con  \cite{Johnson:1999qt} 
found in ${\cal N}=2$ gravity duals - an important part of that analysis 
was understanding that the singularity of the geometry was screened from 
the physics of a D3 brane probe which led to an understanding of how the 
singularity could be removed. It's possible we are seeing hints of
something similar, if more complicated, here, although at this stage we
can not see how to remove the singular behaviour.

\subsection{Large $N$ Goldstone boson ($\eta'$)}

Since there is chiral symmetry breaking via a condensate in the
$m\rightarrow 0$ limit,  we also expect there to be a Goldstone
boson in the meson spectrum. Such a Goldstone mode must exist as a
solution to the DBI equation of motion, as the following
holographic version of the Goldstone theorem shows. Assume a
D7-embedding with $w_5 = 0$ and $w_6 \sim c/\rho^2$
asymptotically.  A small $U(1)$ rotation $\exp(i\epsilon)$ of $w_5
+i w_6$ generates a solution which is a normalizable small
fluctuation about this background, with $w_6$ unchanged (to order
$(\epsilon^2)$) and $w_5 = \epsilon c/\rho^2$. Thus a small
fluctuation with $w_6$ unchanged and $w_5 = \epsilon
\frac{c}{\rho^2} \sin( k \cdot x)$ is a normalizable solution of
the {\it linearized} equations of motion provided $k^2 = 0$.  In
other words there must be a Goldstone boson associated with $w_5$
fluctuations. Note that if the embedding were asymptotically $w_6
\sim m + c(m)/\rho^2$ for non-zero $m$,   a $U(1)$ rotation of
$w_5 + i w_6$ would still generate another solution. However this
solution is no longer a {\it normalizable} small fluctuation about
the original embedding - asymptotically the mass will acquire a 
different phase moving us to a different theory.
Thus if the mass is kept fixed asymptotically one does not find a massless
particle in the spectrum, which reflects the explicit symmetry
breaking by the quark mass $m$.

Note that the $U(1)$ chiral symmetry is non-anomalous only in the
limit $N\rightarrow\infty$.   The Goldstone boson discussed above
is analogous to the $\eta^{\prime}$ in QCD,  which becomes a
Goldstone boson in the large $N$ limit \cite{Wittenetaprime}. Its
masslessness in this setting should not be a surprise, since the
regime of validity of supergravity corresponds to large $N$ in the
dual field theory.  We expect that finite $N$ stringy effects will
give the $\eta^{\prime}$ a non-zero mass.

%The background appears to break chiral symmetries even at $m=0$.
%An obvious signal of chiral symmetry breaking would be a Goldstone
%boson of the U(1)$_A$ symmetry of the quark. This symmetry is the
%U(1) symmetry in the $w_5-w_6$ plane of the geometry. Since $w_6$
%has acquired a condensate like vev the field $w_5$ should play the
%role of the Goldstone field.

In the solutions discussed where $w_6$ has a background value,
fluctuations in $w_5$ should contain the Goldstone mode. Let
us turn to the numerical study of these fluctuations in the background
of the $w_6$ solutions we have obtained above.
The linearized equation of motion for small fluctuations of the
form $w_5 = f(\rho)\sin( k \cdot x)$, with $x$ the four Minkowski
coordinates, are \beq \begin{array}{c} {d
\over d \rho} \left[ {e^{\phi} {\cal G}(\rho,w_6) \over \sqrt{ 1
+ (\partial_\rho w_6)^2}} \partial_\rho f(\rho) \right] + M^2
{e^{ \phi} {\cal G} (\rho,w_6) \over \sqrt{ 1 + (\partial_\rho
w_6)^2}} H \left( { (\rho^2 + w_6^2)^2 + b^4 \over (\rho^2 +
w_6^2)^2 - b^4} \right)^{(1- \delta)/2}  {  (\rho^2 + w_6^2)^2 -
b^4 \over (\rho^2 + w_6^2)^2} f(\rho) \\
\\-\sqrt{ 1 + (\partial_\rho w_6)^2} { 4 b^4 \rho^3 \over (\rho^2 +
w_6^2)^5} \left( { (\rho^2 + w_6^2)^2 + b^4 \over  (\rho^2 +
w_6^2)^2 - b^4} \right)^{\Delta/2} ( 2 b^4 - \Delta (\rho^2 +
w_6^2)^2) f(\rho) = 0 \, . \label{lin}
\end{array}\eeq
The meson mass as a function of quark mass 
for the regular solutions for $w_5$ are plotted in figure 14.
The meson mass indeed falls to zero as the quark mass is taken to zero
providing further evidence of chiral symmetry breaking.

At small $m$,  the mass associated to the $w_5$ fluctuations
scales like $\sqrt{m}$. This is consistent with field theory
expectations.  A well known field theory argument for this scaling
is as follows. The low-energy effective Lagrangian depends on a
field $\eta^{\prime}$ where $\exp(i\eta^{\prime}/f)$ parameterizes
the vacuum manifold and transforms by a phase under chiral $U(1)$
rotations. A quark mass term transforms by the same phase under
$U(1)$ rotations, and thus breaks the $U(1)$ explicitly.  A
chiral Lagrangian consistent with this breaking has a term $\mu^3
Re(m \exp( i \eta^{\prime}/f))$ where $\mu$ is some parameter with
dimensions of mass. For real $m$, expanding this term to quadratic
order gives a mass term $\frac{\mu^3}{f^2}m {\eta^{\prime}}^2$. It
would be very interesting to demonstrate this scaling with $m$
analytically in the DBI/supergravity setting, along with other 
low-energy ``theorems''.

For comparison it is interesting to study 
$w_6$ fluctuations as well,  which we expect to have a mass gap.
Analytically linearizing the $w_6$ equation of motion is
straightforward but the result is unrevealingly messy. Since we
must eventually solve the equation numerically,  we can use a simple
numerical trick to obtain the solutions.  We solve equation
(\ref{eommc}) for $w_6$ above but write $w_6 = w_{6}^0 + \delta
w_6(r)$,  where numerically we enforce $\delta w_6$ to be very
small relative to the background configuration $w_6^0$. 
With this ansatz we retain the field equation in its non-linear
form, but it is numerically equivalent to standard linearization.
%This accounts for the $\rho$ dependence of $\delta w_6$.
We must also add a term to the l.h.s.~of (\ref{eommc})
which takes into account the $x$ dependence of
$\delta w_6$. This dependence
 takes the same form as that in the linearized
$w_5$ equation (\ref{lin}), i.e.~$\delta w_6 =h(\rho)\, {\rm sin}(k
\cdot x)$. The extra term to be added to (\ref{eommc}) is
\beq
\Delta V = M^2 {e^{ \phi} {\cal G}
(\rho,w_6) \over \sqrt{ 1 + (\partial_\rho w_6)^2}}
H \left( {
(\rho^2 + w_6^2)^2 + b^4 \over (\rho^2 + w_6^2)^2 - b^4}
\right)^{(1- \delta)/2}  {  (\rho^2 + w_6^2)^2 - b^4 \over
(\rho^2 + w_6^2)^2}
\delta w_6 \, .
\eeq

The numerical solutions for the $w_6$ fluctuations are  plotted in figure 14.
The  $w_6$ fluctuations have a mass gap, as expected since they 
are transverse to the vacuum manifold.

\begin{figure}[!h]
\begin{center}
\includegraphics[height=7cm,clip=true,keepaspectratio=true]{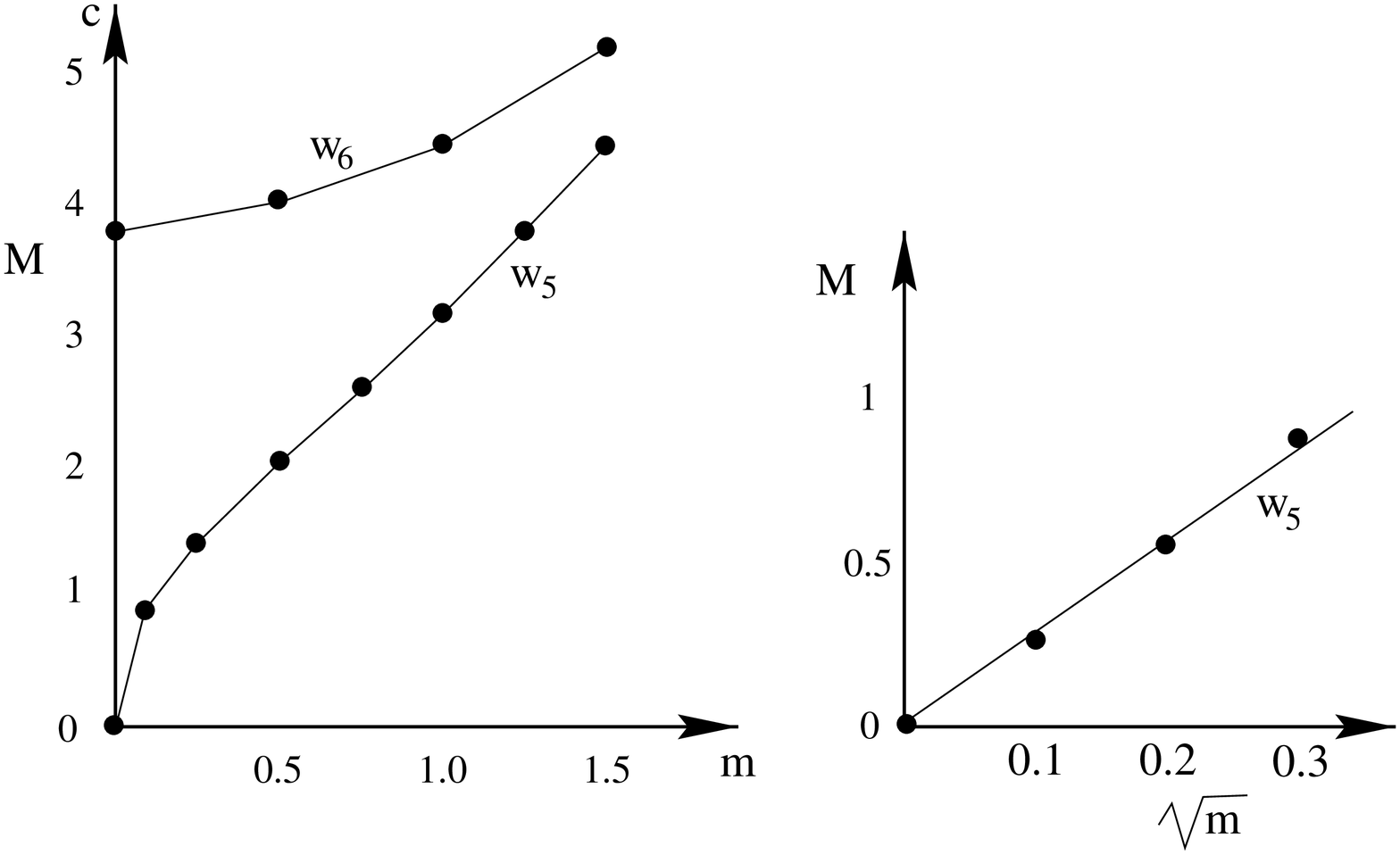}
\caption{A plot of the $w_5$ and $w_6$
meson mass vs quark mass $m$ associated with the fluctuations about
the regular solutions of the equation of motion for the
Constable-Myers flow.The Goldstone mass is also plotted vs $\sqrt{m}$ 
with a linear fit.}
\label{quiv9}
\end{center}
\end{figure}

\subsection{Pions}

Thus far we have found a particle in the spectrum which is morally
equivalent to the $\eta^{\prime}$ in the large $N$ limit of QCD
where it becomes a Goldstone boson.  
In order to obtain true
pions one must have a non-abelian flavour symmetry.
Unfortunately in the background which we consider,  taking 
$N_f >1$ D7-branes does not give rise to a $U(N_f)_L \times U(N_f)_R$ 
chiral symmetry.  Instead one gets only diagonal $U(N_f)$ times an 
axial $U(1)$.  The reason is that the theory contains a coupling 
$\tilde \psi_i X \psi_i$,  where $i$ is a flavor index and $X$ is an 
adjoint scalar without any flavour indices.  The coupling to $X$ explicitly
breaks the  $U(N_f)_L \times U(N_f)_R$ chiral symmetry 
to the diagonal subgroup,  but preserves
an axial $U(1)$ which acts as 
\begin{align}
\psi_i \rightarrow e^{i\theta} 
\psi_i, \qquad \tilde\psi_i \rightarrow e^{i\theta} 
\tilde \psi_i, \qquad X \rightarrow e^{-2i\theta} X.
\end{align}
Thus a $\tilde \psi_i \psi_i$ condensate will only give rise to  
one Goldstone boson, even if 
$N_f > 1$.  If $X$ were massive, there would be an approximate $U(N_f)_L
\times U(N_f)_R$ symmetry at low energies,  but this is not the case in the 
Constable-Myers background. 

Note that $N_f >1$ coincident D7-branes 
may be embedded in the same way as a single
 D7-brane.  In this case there are $N_f$
independent solutions to the linearized equations of motion for
 small fluctuations about this embedding,   
corresponding to fluctuations of the diagonal entries in a diagonal 
$N_f \times N_f$
matrix.  These fluctuations would naively
give rise to at least $N_f$ Goldstone bosons,  rather than one.  
These extra states will not remain massless though since the 
interaction with the scalar fields which breaks the symmetry will 
induce a mass.
% However the 
% flavour symmetry corresponds to a gauge symmetry of the 
% D7-brane action.  Restricting to fluctuations of gauge invariant
% fields such as ${\rm tr}w_5$, where the
% trace is over flavour indices,  there is only one 
% Goldstone boson. 

Nevertheless we can still make a rough comparison 
between our $\eta'$ and QCD 
pions. In a two-flavour large $N$ QCD model where the quarks are 
degenerate one would expect four degenerate Goldstone bosons. As $N$
is decreased, instanton effects will enter to raise the $\eta'$ mass.
However since we are at large $N$, the mass formula for our
Goldstone is that applicable to the pions. 
It is therefore amusing to compare the
Goldstone mass we predict to that of the QCD pion. There is considerable
uncertainty in matching the strong coupling scale of our theory to that of QCD.
In QCD the bare
up or down quark mass is roughly $0.01 \Lambda$ and the pion mass
of order $0.5 \Lambda$ (it is of course hard to know precisely
what value one should pick for $\Lambda$). The comparison to our
theory is a little hard to make but if we assume that $\Lambda
\simeq b = 1$ then for this quark mass we find $m_\pi \simeq 0.25
\Lambda$. The gravity dual is correctly predicting the pion mass
at the level of a factor of two. Of course we cannot  expect a perfect
match given the additional degrees of freedom in the deformed
${\cal N}=4$ theory relative to real QCD. 

\section{Conclusions and open questions}

We have studied two non-supersymmetric gravity backgrounds with
embedded D7-brane probes,  corresponding to Yang-Mills theories with confined
fundamental matter.  The AdS Schwarzschild black-hole
background which in the presence of the probe
describes an ${\cal N}=2$ Yang-Mills theory
at finite temperature,
exhibits interesting behaviour such as a bilinear quark
condensate and a geometric transition which may correspond to a
third order transition in the gauge theory. The D7-brane embedding
into the  Constable-Myers
background is more QCD-like,  showing a chiral
condensate at small quark mass as well as the accompanying pion (or large N 
$\eta^{\prime}$).

%Our analysis has been largely numerical in this paper. It
%would be desirable in the future to distil the
%essential features of the discussion
%analytically. In particular,  understanding 
%the infra-red behaviour
%of the solutions analytically would shed light on
%the screening of the singularity in the Myers-Constable background.

In the Constable-Myers background with a D7-brane, there is a spontaneously
broken $U(1)$ axial symmetry.  A closer approximation to (large N) QCD would  
require a spontaneously broken $U(N_f)_L \times U(N_f)_R$
chiral symmetry with $N_f = 2$ or $N_f =3$.  Unfortunately,  simply adding
D7-branes does not accomplish this in the Constable-Myers background.  
Assuming that one were able to find a background with the full chiral symmetry,
it would be very interesting to obtain a
holographic interpretation of low-energy current algebra
theorems and make predictions for chiral-Lagrangian
parameters based on the non-abelian
Dirac-Born-Infeld action. One challenge would be to obtain such
terms as the Wess-Zumino-Witten term,  $\int\, d^5 x\, {\rm
tr}(\Sigma d \Sigma^{\dagger})^5$. Note that this term requires an
auxiliary fifth dimension, which appears naturally in the
holographic context.

It is clearly important to look at more physical
geometries. We chose these backgrounds because they are
particularly simple; the $S^5$ is left invariant and hence
embedding the D7 is straightforward and the RG flow depends
only on the radial direction. More complicated geometries,
such as the Yang-Mills$^*$ geometry \cite{Babington:2002qt}, 
that include mass
terms for the adjoint scalars and fermions of ${\cal N}=4$,
have extra dependence on the angles of the $S^5$ and the
resulting equations of motion are much less tractable. 

In conclusion the results presented here represent
another success for the AdS/CFT Correspondence. The
results suggest that gravity duals of non-supersymmetric
gauge theories may induce chiral symmetry breaking if light
quarks are introduced, just as is observed in QCD. This
opens up the possibility of studying the light
meson sector of QCD using these new techinques.

%%%%%%%%%%%%%%%%%%
%\vspace{1,5cm}

\newpage

{\bf Acknowledgements}

We are very grateful to R.~Brower, N.~Constable, A.~Hanany,
C.~N\'{u}\~{n}ez, M. ~Petrini and
N.~Prezas for enlightening discussions.

The research of J.E., Z.G.~and I.K.~is supported by DFG (Deutsche
Forschungsgemeinschaft) within the `Emmy Noether' programme, grant
ER301/1-2. J.B.~acknowledges support through a Research Fellowship of
the Alexander von Humboldt Foundation.  
N.E. is grateful for the support of a PPARC Advanced Research Fellowship.

\end{document}